\documentclass[prd,aps,floats,amssymb,preprint,nofootinbib]{revtex4}

\usepackage{graphicx}

\usepackage{epsfig}
\usepackage{amssymb}
\usepackage{bbm}
\usepackage{color}
\usepackage{ulem}

\usepackage{enumitem}
\usepackage{graphicx}
\usepackage{subfigure}
\usepackage{amsmath}
\usepackage{bbm}
\usepackage{color}

\newcommand{\be}{\begin{equation}}
\newcommand{\ee}{\end{equation}}
\newcommand{\bea}{\begin{eqnarray}}
\newcommand{\eea}{\end{eqnarray}}
\newcommand{\beas}{\begin{eqnarray*}}
\newcommand{\eeas}{\end{eqnarray*}}

\newcommand\mpl{M_{\rm Pl}}

\begin{document}

\title{Cosmic acceleration and the challenge of modifying gravity} 

\author{Mark Trodden}

\affiliation{Center for Particle Cosmology, Department of Physics and Astronomy, University of Pennsylvania,
Philadelphia, Pennsylvania 19104, USA}

\footnote{trodden@physics.upenn.edu}

\begin{abstract}
I briefly discuss the challenges presented by attempting to modify general relativity to obtain an explanation for
the observed accelerated expansion of the universe. Foremost among these are the questions of theoretical consistency - the
avoidance of ghosts in particular - and the constraints imposed by precision local tests of gravity within the solar system. For those
models that clear these highly constraining hurdles, modern observational cosmology offers its own suite of tests, improving with
upcoming datasets, that offer the possibility of ruling out modified gravity approaches or providing an intriguing hint of new
infrared physics. In the second half of the talk, I discuss a recent approach to extracting cosmology from higher-dimensional
induced gravity models.
\end{abstract}

\maketitle

\setcounter{footnote}{0}

%%%%%%%%%%%%%%%%%%

\section{Introduction}
The discovery of late-time cosmic acceleration has led cosmologists to carefully examine the possible contributions to the mass-energy of the universe
that might source this behavior within general relativity (GR). Perhaps more provocatively, a second possibility has also been considered, namely
that GR itself may not provide the correct set of rules with which to understand how the known matter and radiation content affects
the universe on the largest scales. It may be that curvatures and length scales in the observable universe are only now reaching values at which an infrared modification of gravity can make itself apparent by driving self-acceleration (for reviews see~\cite{Copeland:2006wr,Frieman:2008sn,Silvestri:2009hh,Caldwell:2009ix}).

General relativity is very well tested in the solar system, in measurements of the
period of the binary pulsar, and in the early universe, via primordial nucleosynthesis.
None of these tests, however, probes the ultra-large length scales and
low curvatures characteristic of the Hubble radius today.  It is therefore
{\it a priori} conceivable that gravity is modified in the very far
infrared, in such a way that the universe begins to accelerate at late
times.

In practice, however, as I will describe below, it is difficult to construct a simple model that
embodies this hope.  A straightforward possibility is to modify the usual
Einstein-Hilbert action by adding new covariant terms constructed from the scalar invariants of the theory.
Such theories can lead to late-time acceleration, but unfortunately typically lead to one of two problems. Either they 
are in conflict with tests of GR in the solar system, due to the existence of additional 
dynamical degrees of freedom, or they contain ghost-like degrees of freedom
that seem difficult to reconcile with fundamental theories.  Nevertheless, a restricted class of such theories remain viable, and
should be further constrained by upcoming cosmological missions.

A more dramatic strategy is to imagine that we live on a brane embedded
in a large extra dimension.  Although such theories can lead to perfectly
conventional gravity on large scales, it is also possible to choose the
dynamics in such a way that new effects show up exclusively in the far
infrared.  Such theories can naturally
lead to late-time acceleration, but may 
have strong-coupling or ghost issues.  Nevertheless, these models hold out
the possibility of having interesting and testable predictions that
distinguish them from models of dynamical dark energy. 

In this talk, delivered at the 28th International Colloquium on Group Theoretical Methods in Physics (ICGTMP), I attempted to provide an overview of theoretical approaches to this problem, and to describe some of the challenges, both theoretical and observational, faced by attempts to address cosmic acceleration in this way. The presentation was intended to explain the key ideas and to highlight some very recent work of my own. For this reason, and because this writeup is necessarily brief, I have chosen to reference only selected review articles, those papers to which I directly referred in the talk itself, and several that came out directly afterwards..

\section{Modifying Gravity}
Although, within the context of General Relativity (GR), one doesn't think about it too often, the metric tensor contains, in principle, more degrees of freedom than the usual spin-2 {\it graviton}.
The reason why one doesn't hear of these degrees of freedom in GR is that the Einstein-Hilbert action is a very special choice, resulting in second-order equations of motion, which constrain away the scalars and the vectors, so that they are non-propagating. However, this is not the case if one departs from the Einstein-Hilbert form for the action. When using any modified action (and the usual variational principle) one inevitably frees up some of the additional degrees of freedom. In fact, this can be a good thing, in that the dynamics of these new degrees of freedom may be precisely what one needs to drive the accelerated expansion of the universe. However, there is often a price to pay.

The problems may be of several different kinds. First, there is the possibility that along with the desired deviations from GR on cosmological scales, one may also find similar deviations on solar system scales, at which GR is rather well-tested. Second is the possibility that the newly-activated degrees of freedom may be badly behaved in one way or another; either having the wrong sign kinetic terms (ghosts), and hence being unstable, or leading to superluminal propagation, which may lead to other problems.

These constraints are surprisingly restrictive when one tries to create viable modified gravity models yielding cosmic acceleration. In the next few sections I will describe several ways in which one might modify the action, and in each case demonstrate how cosmic acceleration emerges. However, I will also point out how the constraints I have mentioned rule out these simple examples, and mention how one must complicate the models to recover viable models.

\subsection{A Simple Model: $f(R)$ Gravity}
The simplest way one could think to modify GR to obtain cosmic acceleration is to replace the Einstein-Hilbert Lagrangian density by a general function $f(R)$ of the Ricci scalar $R$~\cite{Carroll:2003wy,Capozziello:2003tk}.
\be
S=\frac{\mpl^2}{2}\int d^4 x\sqrt{-g}\, \left[R+f(R)\right] + \int d^4 x\sqrt{-g}\, {\cal L}_{\rm m}[\chi_i,g_{\mu\nu}] \ ,
\label{jordanaction}
\ee
where $\mpl\equiv (8\pi G)^{-1/2}$ is the (reduced) Planck mass and ${\cal L}_{\rm m}$ is the Lagrangian
density for the matter fields $\chi_i$.

Here, I have written the matter Lagrangian as ${\cal L}_{\rm m}[\chi_i,g_{\mu\nu}]$ to make explicit that in
this frame - the {\it Jordan} frame - matter falls along geodesics of the metric $g_{\mu\nu}$.

The equation of motion obtained by varying the action~(\ref{jordanaction}) is
\be
\left(1+f_R \right)R_{\mu\nu} - \frac{1}{2}g_{\mu\nu}\left(R+f\right)
+ \left(g_{\mu\nu}\nabla^2 -\nabla_\mu\nabla_\nu\right) f_R
=\frac{T_{\mu\nu}}{\mpl^2} \ ,
\label{jordaneom}
\ee
where I have defined $f_R\equiv \partial f/\partial R$.

Further, if the matter content is described as a perfect fluid, with energy-momentum tensor,
\begin{equation}
T_{\mu\nu}^m = (\rho_m + p_m)U_{\mu} U_{\nu} + p_m g_{\mu\nu}\ ,
\label{perfectfluid}
\end{equation} 
where $U^{\mu}$ is the fluid rest-frame four-velocity, $\rho_m$ is the energy density and $p_m$ is the pressure, then the fluid equation of motion is the usual continuity equation. 

When considering the background cosmological evolution of such models, the metric can be taken as the flat Robertson-Walker form, $ds^2=-dt^2+a^2(t)d{\bf x}^2$. In this case, the usual Friedmann equation of GR is modified to become
\be
3H^2 -3f_R ({\dot H}+H^2)+\frac{1}{2}f+18f_{RR}H({\ddot H}+4H{\dot H})=\frac{\rho_m}{\mpl^2}
\label{jordanfriedmann}
\ee
and the continuity equation is
\be
{\dot \rho}_m +3H(\rho_m+p_m)=0 \ .
\label{jordancontinuity}
\ee
When supplied with an equation of state parameter $w$, the above equations are sufficient to solve for the
background cosmological behavior of the space-time and it's matter contents. For appropriate choices of the
function $f(R)$ it is possible to obtain late-time cosmic acceleration without the need for dark energy, 
although evading bounds from precision solar-system tests of gravity turns out to be a much trickier
matter, as we shall see.

While one can go ahead and analyze this theory in the Jordan frame, it is more convenient to perform a carefully-chosen conformal transformation on the metric, in order to render the gravitational action in the usual Einstein Hilbert form of GR. Consider writing
\be
{\tilde g}_{\mu\nu} = \Omega(x^{\alpha}) g_{\mu\nu} \ ,
\label{conftrans}
\ee
and construct the function $r(\Omega)$ that satisfies
\be
1+f_R[r(\Omega)]=\Omega \ .
\ee
Defining a rescaled scalar field by $\Omega \equiv e^{\beta\phi}$, with 
$\beta\mpl\equiv\sqrt{2/3}$, the resulting action becomes
\bea
{\tilde S}=\frac{\mpl}{2}\int d^4 x\sqrt{-{\tilde g}}\, {\tilde R} &+&\int d^4 x\sqrt{-{\tilde g}}\, 
\left[-\frac{1}{2}{\tilde g}^{\mu\nu}(\partial_{\mu}\phi)\partial_{\nu}\phi -V(\phi)\right] \nonumber \\ 
&+&
\int d^4 x\sqrt{-{\tilde g}}\, e^{-2\beta\phi} {\cal L}_{\rm m}[\chi_i,e^{-\beta\phi}{\tilde g}_{\mu\nu}]\ ,
\label{einsteinaction}
\eea
where the potential $V(\phi)$ is determined entirely by the original form~(\ref{jordanaction}) 
of the action and is given by
\be
V(\phi)=\frac{e^{-2\beta\phi}}{2}\left\{e^{\beta\phi}r[\Omega(\phi)] - f(r[\Omega(\phi)]) \right\} \ .
\label{einsteinpotential}
\ee

The equations of motion in the Einstein frame are much more familiar than those in the Jordan frame, 
although there are some crucial subtleties. In particular, note that in general, test particles of the matter content $\chi_i$ do not freely fall along geodesics of the metric ${\tilde g}_{\mu\nu}$.

The equations of motion in this frame are those obtained by varying the action with respect to the metric ${\tilde g}_{\mu\nu}$
\be
{\tilde G}_{\mu\nu} = \frac{1}{\mpl^2}\left({\tilde T}_{\mu\nu} + T^{(\phi)}_{\mu\nu}\right) \ ,
\label{einsteineom}
\ee
with respect to the scalar field $\phi$
\be
{\tilde \nabla^2}\phi = -\frac{dV}{d\phi}(\phi) \ ,
\label{scalareom}
\ee
and with respect to the matter fields $\chi_i$, described as a perfect fluid.

Once again, I will specialize to consider background cosmological evolution in this frame. The 
Einstein-frame line element can be written in familiar FRW form as
\be
ds^2 =-d{\tilde t}^2+{\tilde a}^2({\tilde t})d{\bf x}^2 \ ,
\label{einsteinFRWmetric}
\ee
where $d{\tilde t}\equiv\sqrt{\Omega}\, dt$ and ${\tilde a}(t)\equiv\sqrt{\Omega} \,a(t)$. The Einstein-frame matter energy-momentum tensor is then given by
\be
{\tilde T}_{\mu\nu}^m = ({\tilde \rho}_m + {\tilde p}_m){\tilde U}_{\mu} {\tilde U}_{\nu} + 
{\tilde p}_m {\tilde g}_{\mu\nu}\ ,
\label{einsteinperfectfluid}
\ee
where ${\tilde U}_{\mu}\equiv \sqrt{\Omega} \,U_{\mu}$, ${\tilde \rho}_m\equiv \rho_m/\Omega^2$ and 
${\tilde p}_m\equiv p_m/\Omega^2$.

Now, as I mentioned in the introduction, any modification of the Einstein-Hilbert action must, of
course, be consistent with the classic solar system tests of gravity
theory, as well as numerous other astrophysical dynamical tests. 
We have chosen the coupling constant $\mu$ to be very small, but
we have also introduced a new light degree of freedom.  As shown by
Chiba~\cite{Chiba:2003ir}, the simple model above is equivalent to a Brans-Dicke
theory with $\omega=0$ in the approximation where the potential
was neglected, and would therefore be inconsistent with solar system measurements~\cite{Bertotti:2003rm}.

To construct a realistic $f(R)$ model requires a more complicated function, with more than one
adjustable parameter in order to fit the cosmological data~\cite{Bean:2006up} and satisfy solar system bounds. 

\subsection{Extensions: Higher-Order Curvature Invariants}
It is natural to consider generalizing the action of~\cite{Carroll:2003wy} to include other curvature invariants~\cite{Carroll:2004de}.
There are, of course, any number of terms that one could consider, but for simplicity, focus on
those invariants of lowest mass dimension that are also parity-conserving $P \equiv  R_{\mu\nu}\,R^{\mu\nu}$ and 
$Q \equiv  R_{\alpha\beta\gamma\delta}\,R^{\alpha\beta\gamma\delta}$.

The action then takes the form
\begin{equation}
S=\int d^4x \sqrt{-g}\,[R+f(R,P,Q)] +\int d^4 x\, \sqrt{-g}\,
{\cal L}_M \ ,
\label{genaction}
\end{equation}
where $f(R,P,Q)$ is a general function describing deviations from general relativity.

Actions of the form~(\ref{genaction}) generically admit a maximally-symmetric solution that is often unstable to another accelerating power-law attractor. 
It has been shown that solar system constraints, of the type I have described for $f(R)$ models, can be evaded by these more general 
models when, for example, the $Q$ terms are relevant on those scales. However, these theories generically contain ghosts and/or superluminally propagating modes~\cite{Chiba:2005nz,Navarro:2005da,DeFelice:2006pg,Calcagni:2006ye}. I therefore will not discuss them further here.

\subsection{Induced Gravity Models}

In the Dvali-Gabadadze-Porrati (DGP) model~\cite{Dvali:2000hr}, our observed $4D$ universe is embedded in an infinite empty fifth dimension. 
Despite the fact that the extra dimension is infinite in extent, the inverse-square law is nevertheless recovered at short distances on the brane due to an intrinsic, four-dimensional Einstein-Hilbert term in the action
\be
	S_{\rm DGP} = \int_{\rm bulk} {\rm d}^5x\sqrt{-g_5}\frac{M_5^3}{2}R_5 +
\int_{\rm brane} {\rm d}^4x \sqrt{-g_4} \left(\frac{M_4^2}{2}R_4 + {\cal L}_{\rm matter}\right)\,.
\ee
The Newtonian potential on the brane scales as $1/r$ at short distances, as in $4D$ gravity, and asymptotes to $1/r^2$ at large distances, characteristic of $5D$ gravity. The cross-over scale $m_5^{-1}$ between these two behaviors is set by the bulk and brane Planck masses ($M_{5}$ and $M_{4}$ respectively) via $m_5 = \frac{M_5^3}{M_4^2}$.

In this picture, the higher-dimensional nature of gravity affects the $4D$ brane through deviations from general relativity on horizon scales, that may give rise to the observed accelerated expansion. This model faces its own challenges however. The branch of solutions that include self-acceleration suffers from ghost-like instabilities, and on the observational front, DGP cosmology is statistically disfavored in comparison to $\Lambda$CDM and is significantly discordant with constraints on the curvature of the universe. 

\section{Cascading Cosmology}

These facts, among others, have led to the idea of {\it cascading gravity}~\cite{deRham:2007xp,deRham:2007rw,Kaloper:2007ap} --- a higher-dimensional generalization of the DGP idea, which is free of divergent propagators and ghost instabilities. In this model one embeds a succession of higher-codimension branes into each other, with energy-momentum confined to the $4D$ brane and gravity living in higher-dimensional space. An important first test, which I've been working on recently~\cite{Agarwal:2009gy}, is whether such models can reproduce a successful cosmological evolution. 

\subsection{A Proxy Theory for Cascading Gravity}

The main idea is to embed a 3-brane in a succession of higher-dimensional DGP branes, each with their own Einstein-Hilbert term. Denote coordinates in the full six dimensional spacetime by $x^{0},x^{1},x^{2},x^{3},x^{5},x^{6}$. Indices $M,N,...$ run over 0,1,2,3,5 (i.e. the $4+1D$ coordinates), indices $\mu,\nu,...$ run over 0,1,2,3 (i.e. the $3+1D$ coordinates), and indices $i,j,...$ run over $1,2,3$ (i.e. the $3D$ spatial coordinates). Further denote the fifth and sixth dimensional coordinates by $y=x^{5}$ and $z=x^{6}$, where convenient. The action is then
\bea
\nonumber
	S_{\rm cascade} & = & \int_{\rm bulk} {\rm d}^6x\sqrt{-g_6}\frac{M_6^4}{2}R_6 + \int_{\rm 4-brane} {\rm d}^5x\sqrt{-g_5}\frac{M_5^3}{2}R_5 \\
	& & + \int_{\rm 3-brane} {\rm d}^4x \sqrt{-g_4} \left(\frac{M_4^2}{2}R_4 + {\cal L}_{\rm matter}\right)\,.
\label{S6}
\eea
As a result, the force law on the 3-brane ``cascades" from $1/r^2$ to $1/r^3$ to $1/r^4$ as one moves increasingly far from a source, with the $4D\rightarrow 5D$ and $5D\rightarrow 6D$ cross-over scales given respectively by $m_5^{-1}$ and $m_6^{-1}$, with $m_6 = \frac{M_6^4}{M_5^3}$

The next question is, of course, whether the resulting cosmology is consistent with current observations, and whether it offers distinguishing signatures from $\Lambda$CDM cosmology. Unfortunately, finding analytical solutions is a significant challenge, even in the simplest $6D$ case, as the bulk metric is generally expected to depend on all extra-dimensional coordinates plus time~\cite{Chatillon:2006vw}.

To proceed, consider, in analogy with a useful decoupling limit for the DGP model, the limit $M_5,M_6\rightarrow \infty$, with the strong-coupling scale $\Lambda_6 = (m_6^4M_5^3)^{1/7}$
kept fixed. In this limit, the action~(\ref{S6}) may be expanded around flat space, and reduces to a local theory on the 4-brane, describing $5D$ weak-field metric perturbations $h_{MN}$ and an interacting scalar field $\pi$. The resulting action is~\cite{deRham:2007xp}
\begin{eqnarray}
	S_{\rm decouple} & = & \frac{M_{5}^{3}}{2} \int_{\rm bulk} {\rm d}^{5}x \left[ -\frac{1}{2} h^{MN}(\mathcal{E}h)_{MN} + \pi\eta^{MN}(\mathcal{E}h)_{MN} - \frac{27}{16m_6^2} (\partial\pi)^{2} \Box_{5}\pi \right] \nonumber \\
	& & + \int_{\rm brane} {\rm d}^{4}x \left[ -\frac{M_{4}^{2}}{4} h^{\mu\nu}(\mathcal{E}h)_{\mu\nu} + \frac{1}{2} h^{\mu\nu}T_{\mu\nu} \right]\,,
\label{5dcov1}
\end{eqnarray}
where $(\mathcal{E}h)_{MN}$ is the linearized Einstein tensor in $5D$, and $(\mathcal{E}h)_{\mu\nu}$ that in $4D$. Nearly all of the interesting features of DGP gravity are due to the helicity-0 mode $\pi$ and can be understood at the level of the decoupling theory. 

Of course~(\ref{5dcov1}) is restricted to weak-field gravity and therefore cannot be used to find cosmological solutions. As a ``proxy" brane-world scenario, we proposed to complete~(\ref{5dcov1}) into a covariant, non-linear theory of gravity in $5D$ coupled to a 3-brane, using 
\bea
\nonumber
	S & = & \frac{M_{5}^{3}}{2}\int_{\rm bulk}{  {\rm d}^{5}x\sqrt{-g_{5}}\left[e^{-3\pi/2}R_{5} - \frac{27}{16m_6^2}(\partial\pi)^2 \Box_{5}\pi \right]} \\
	& & + \int_{\rm brane}{{\rm d}^{4}x\sqrt{-g_{4}}\left[ \frac{M_{4}^{2}}{2}R_{4} + \mathcal{L}_{\rm{matter}} \right]}\,.
\label{5dcov}
\eea
By construction this theory reduces to~(\ref{5dcov1}) in the weak-field limit, and therefore agrees with cascading gravity to leading order in $1/M_5$. The proposed $5D$ completion is by no means unique, but the hope is that the salient features of cascading cosmology are captured by the $5D$ effective theory, and that the resulting predictions are at least qualitatively robust to generalizations of~(\ref{5dcov}).

\subsection{Covariant Equations of Motion On and Off the Brane}
\label{eoms}
%%%%%%%%%%%%%%%%%%%%%%%%%%%%%%%%%%%%%%%%%%%%%%%%%%%%%%%%%%%%

The bulk Einstein equations are
\bea
\nonumber
	e^{-3\pi/2} G_{MN} & = & -\frac{27}{16m_6^2}\left[ \partial_{(M}(\partial\pi)^{2}\partial_{N)}\pi  - \frac{1}{2}g_{MN}\partial^{K}(\partial\pi)^{2}\partial_{K}\pi - \partial_{M}\pi\partial_{N}\pi\Box_{5}\pi \right] \\
	& & - \left( g_{MN}\Box_{5} - \nabla_{M}\nabla_{N} \right) e^{-3\pi/2} \ ,
\label{bulkein}
\eea
where $G_{MN}$ is the $5D$ Einstein tensor, and the $\pi$ equation of motion is 
\begin{eqnarray}
	(\Box_{5}\pi)^{2} - (\nabla_{M}\partial_{N}\pi)^{2} - R^{MN} \partial_{M}\pi \partial_{N}\pi  = \frac{4}{9}m_6^2e^{-3\pi/2}R_{5},
\label{pieom1}
\end{eqnarray}
where $R_{MN}$ is the $5D$ Ricci tensor and $R_{5}$ is the Ricci scalar. Remarkably, even though the cubic $\pi$ interaction has four derivatives, all higher-derivative terms cancel in the variation, yielding a second-order equation of motion for $\pi$.  This is typical of a larger class of ``galileon" theories~\cite{Nicolis:2008in}. I won't have time to speak in detail about these here, but would like to highlight that we are close to completing an 
extension of these theories to the multi-field case, which will allow the exploration of the low-energy limit of braneworld models in co-dimension greater than one\footnote{This work~\cite{Hinterbichler:2010xn} (see also~\cite{Deffayet:2009mn,Deffayet:2010zh,Padilla:2010de,Padilla:2010tj}), and some initial explorations~\cite{Andrews:2010km,Goon:2010xh}, were recently completed.}.

We also require the Israel junction condition
\begin{eqnarray}
\nonumber
	2M_{5}^{3} e^{-3\pi/2} \left( K q_{\mu\nu} - K_{\mu\nu} - \frac{3}{2} q_{\mu\nu} {\cal L}_{n} \pi \right) & = & \frac{27}{8}\frac{M_{5}^{3}}{m_{6}^{2}} \left( \partial_{\mu}\pi \partial_{\nu}\pi {\cal L}_{n} \pi + \frac{1}{3} q_{\mu\nu} \left( {\cal L}_{n}\pi \right)^{3} \right) \\
	& & + \ T^{(4)}_{\mu\nu} - M_4^2G_{\mu\nu}^{(4)} \,,
	\label{covjc1}
\end{eqnarray}
where
\be
	T_{\mu\nu}^{(4)} \equiv - \frac{2}{\sqrt{-q}} \frac{\delta (\sqrt{-q}{\cal L}_{\rm matter})}{\delta q^{\mu\nu}}
\ee
is the matter stress-energy tensor on the brane, and $G_{\mu\nu}^{(4)}$ is the Einstein tensor derived from the induced metric $q_{\mu\nu}$. Similarly,  the boundary condition for $\pi$ on the brane is
\be
e^{-3\pi/2}K + \frac{9}{8m_{6}^{2}} \Big( K_{\mu\nu} \partial^{\mu}\pi \partial^{\nu}\pi + 2{\cal L}_{n}\pi \Box_{4}\pi + K ({\cal L}_{n} \pi)^{2} \Big) = 0\,.
\label{covjc2}
\ee
(Note that equations~(\ref{covjc1}) and~(\ref{covjc2}) are not independent, of course; the divergence of~(\ref{covjc1}) can be shown to be proportional to~(\ref{covjc2}) after using the bulk momentum constraint equation.)
%cccccccccccccccccccccccccccccccccccccccccccccccccccccccccccccccccccccccccccccccccccccccccccccccccccccccccccccccccc

%%%%%%%%%%%%%%%%%%%%%%%%%%%%%%%%%%%%%%%%%%%%%%%%%
\subsection{The Cosmological Evolution on the Brane}
\label{cosmology}
%%%%%%%%%%%%%%%%%%%%%%%%%%%%%%%%%%%%%%%%%%%%%%%%%

The study of brane-world cosmology requires using the equations of motion to obtain a Friedmann equation on the brane, assuming homogeneity and isotropy along the 3+1 world-volume dimensions. The junction conditions~(\ref{covjc1}) and~(\ref{covjc2}) do not form a closed system of equations for $q_{\mu\nu}$, hence deriving an induced Friedmann equation requires knowledge of the bulk geometry~\cite{Chatillon:2006vw}. 

Because of the bulk scalar field, there is no Birkhoff theorem to ensure that the bulk solutions are necessarily static under the assumption of homogeneity and isotropy on the brane --- the most general bulk geometry depends on both the extra-dimensional coordinate {\it and} time. For concreteness, however, focus here on a static warped geometry with Poincar\'e-invariant slices, representing a tractable first step for which, as we will see, the resulting phenomenology is already surprisingly rich.

Writing
\be
{\rm d}s^2_{\rm bulk} = a^2(y) (-{\rm d}\tau^2 + {\rm d}\vec{x}^2) + {\rm d}y^2\, ,
\label{staticbulk}
\ee
the brane motion is governed by two functions, $y(t)$ and $\tau(t)$, describing the embedding, where $t$ is proper time on the brane. The induced metric is of the Friedmann-Robertson-Walker (FRW) form, with spatially-flat ($k=0$) constant-time hypersurfaces ${\rm d}s^{2}_{\rm brane} = -{\rm d}t^{2} + a^{2}(y){\rm d}\vec{x}^2$, where, by virtue of $t$ being the proper time,
\begin{eqnarray}
	\left(\frac{{\rm d} t}{{\rm d}\tau}\right)^2 = a^2 - \left(\frac{{\rm d}y}{{\rm d}\tau}\right)^2\,.
\label{dtaudt}
\end{eqnarray}
Given a solution $a(y)$ to the bulk equations~(\ref{bulkein})--(\ref{pieom1}), the covariant junction conditions (\ref{covjc1}) and (\ref{covjc2})
allow us to solve for the embedding $(y(t),\tau(t))$, and hence the cosmology induced by brane motion through the warped bulk.

For the stress energy on the brane, we assume a collection of (non-interacting) perfect fluids with energy densities $\rho^{(i)}_{\rm m}$ and pressures $P^{(i)}_{\rm m}$, obeying the standard continuity equations. To derive the Friedmann equation on the brane, we use the $(0,0)$ component of~(\ref{covjc1}). Since $\partial_{0}\pi = \pi' {\rm d}y/{\rm d}t$ and ${\rm d}y/{\rm d}t = aH/a'$, we can write the resulting equation as the standard Friedmann equation with an additional effective energy density $\rho_{\pi}$ resulting from the $\pi$ field,
\begin{eqnarray}
3H^2M_4^2 &=& \sum_i \rho^{(i)}_{\rm m} + \rho_\pi\,,
\label{staticjc}
\eea
where 
\bea
\rho_{\pi}&\equiv &M_5^3 \sqrt{a'^{2} + a^{2}H^{2}}  \left\{\frac{9}{8m_6^2} \left( 2\left(\frac{aH}{a'}\right)^{2} - 1 \right) \frac{\pi'^{3}}{a'}-6 e^{-3\pi/2} \left( \frac{\pi'}{2a'} - \frac{1}{a} \right) \right\} \,,
\label{rhopi}
\eea
encoding all the complexity and new physics of the cascading cosmology model. Given a solution $a(y)$, $\pi(y)$ to the bulk equations, this relation
may be inverted to obtain $y(a)$, and used to express all $y$-dependent terms in $\rho_\pi$ as functions of $a$. Equation~(\ref{staticjc}), together with the continuity equations for the brane matter, then forms a closed system for the brane scale factor $a(t)$. 

%%%%%%%%%%%%
\subsection{Numerical Solutions}
\label{numevol}
%%%%%%%%%%%% 

%===========================FIGURE 2 ==================== 
\begin{figure}[!t]
  \begin{center}
    \includegraphics[width=3.5in,angle=0]{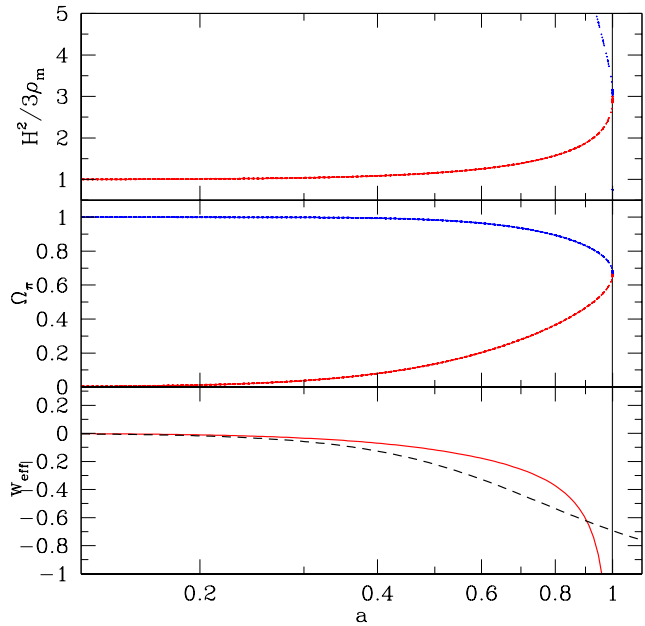}
    \caption{See text for explanation} \label{fig2}
  \end{center}
\end{figure}
%========================================================

To get a feel for the type of cosmological solutions that this model may exhibit requires numerically evolving the full bulk and brane equations in the presence of matter on the brane. To be explicit, assume zero spatial curvature on the brane, include relativistic and pressureless components consistent with the standard cosmological model: $\Omega_{\rm m}=0.3$, $\Omega_{\rm r} =8.5\times 10^{-4}$, and fix the scale factor today to be $a_0=1$. 

%%%%
In the figure, example evolution histories in which no cosmological constant is present to drive cosmic acceleration are plotted. The top panel shows the deviation of the expansion history from that derived from standard matter (for which $3H^2/\rho_m=1$). The curves each show consistent solutions to the modified Friedmann equation (\ref{staticjc}): one solution (red, thick line) recovers the standard expansion history at early times and then undergoes accelerated expansion at late times; the other solution (blue, dotted line) has an expansion history entirely inconsistent with that of standard $\Lambda$CDM, with the $\pi$ field dominating the expansion at all eras, and undergoing heavily decelerated expansion at late times. 

The center panel shows the evolution of the effective fractional energy density, $\Omega_{\pi}=8\pi G\rho_\pi/3H^2$, for these two solutions. For the accelerating solution, the phantom-like behavior in the matter era allows the $\pi$ field to dominate and drive cosmic acceleration at late times. The model is not physical, however, since as $\Omega_\pi\rightarrow 2/3$ one finds $\dot{H}\rightarrow \infty$ and a singularity occurs. This singularity is of an unusual nature - the bulk geometry is smooth and it is the brane embedding that is singular. It is possible therefore that this singularity could be circumvented by the use of a more general metric ansatz, but this has yet to be demonstrated.

Finally, the bottom panel shows a comparison of the effective equation of state for the expansion, $w_{\rm eff} = -1-(2/3){\rm d}\ln H/{\rm d}\ln a$, for the accelerating $\pi$ (red, full line) and fiducial $\Lambda$CDM (black, dashed line) scenarios. The $\pi$ driven expansion histories assume the numerical values $H_{0} = 2.33 \times 10^{-4}$ Mpc$^{-1}$, $m_{6} = 3.5\times 10^{-18}$ Mpc$^{-1}$ and $m_{5} = 4.4\times 10^{-31} $Mpc$^{-1}$ for which the maximum singularity occurs just after $a=1$.
%%%%%

\section{Conclusion} 

Among the possible explanations for the observed accelerated expansion of the universe, the possibility that general relativity may become modified on the largest scales is a particularly intriguing one. In this talk I have outlined a number of modern approaches to this problem, focusing, as expected, on those that I have been involved with in one way or another. I have described how the combined constraints of theoretical consistency, solar system measurements, and cosmological observations tightly bound the possible viable models. I have also discussed very recent work on cascading cosmology, and have shown that cosmic acceleration can arise within this model without the need for a cosmological constant. However, for the limited solutions explored thus far, one hits a singularity in the expansion history so that the universe cannot smoothly transition towards $\Omega_\pi\rightarrow 1$

As I mentioned in the talk, out of the higher dimensional constructions, such as the DGP model, an interesting set of four dimensional effective field theories - the {\it galileons} - arises, encapsulating the effects of modifying gravity. The work which was underway, generalizing that work to multi-galileon theories, is now completed, and we have been hard at work on constraining the resulting Lagrangians. 
Now as then, much remains to be done to understand the full range of acceptable models.

\acknowledgments
I would like to thank the organizers of the ICGTMP-2010 conference, and in particular Wojciech Zakrzewski, for inviting me to deliver this talk.  I would also like to thank my collaborators, from whose joint work with me I have borrowed liberally in putting together this summary. This work is supported in part by NASA ATP grant NNX08AH27G, NSF grant PHY-0930521, Department of Energy grant DE-FG05-95ER40893-A020, and by the Fay R. and Eugene L. Langberg chair.

\end{document}